\documentclass[aps,pre,superscriptaddress,longbibliography]{revtex4-2}

\usepackage{hyperref}
\usepackage{nameref}
\usepackage{amsmath,amssymb}
\usepackage{graphicx}
\usepackage{dcolumn}
\usepackage{bm}
\usepackage[dvipsnames]{xcolor}

\graphicspath{ {Figures/}{ActualFigures/} }

\begin{document}

\title{Nonequilibrium statistics of barrier crossings with competing pathways}
\author{Gulzar Ahmad}
\affiliation{Higher Education Department, Govt.~of the Punjab, Lahore, 54000, Pakistan}
\affiliation{Department of Mathematical Sciences and Interdisciplinary Centre for Mathematical Modelling, Loughborough University, Loughborough, LE11 3TU, United Kingdom}
\author{Sergey Savel’ev}
\affiliation{%
 Department of Physics, Loughborough University, Loughborough, LE11 3TU, United Kingdom}
\author{Steven P.~Fitzgerald}
\affiliation{Department of Applied Mathematics, University of Leeds, Leeds LS2 9JT, United Kingdom}
\author{Marco G.~Mazza}
\affiliation{Department of Mathematical Sciences and Interdisciplinary Centre for Mathematical Modelling, Loughborough University, Loughborough, LE11 3TU, United Kingdom}
\author{Andrew J.~Archer}
\affiliation{Department of Mathematical Sciences and Interdisciplinary Centre for Mathematical Modelling, Loughborough University, Loughborough, LE11 3TU, United Kingdom}

\date{\today}

\begin{abstract}
Many biological, chemical, and physical systems are underpinned by stochastic transitions between equilibrium states in a potential energy.
Here, we consider such transitions in a minimal model with two possible competing pathways, both starting from a local potential energy minimum and eventually finding  the global minimum.
There is competition between the distance to travel in state space and the height of the potential energy barriers to be surmounted, for the transition to occur.
One pathway has a higher energy barrier to go over, but requires traversing a shorter distance, whereas the other pathway has a lower potential barrier but it is substantially further away in configuration space. 
The most likely pathway taken depends on the available time for the transition process; when only a relatively short time is available, the most likely path is the one over the higher barrier. 
We find that upon varying temperature the overall most likely pathway can switch from one to the other.
We calculate the statistics of where the barrier crossing occurs and the distribution of times taken to reach the potential minimum.
Interestingly, while the configuration space statistics is complex, the time of arrival statistics is rather simple, having an exponential probability density over most of the time range.  Taken together, our results show that empirically observed rates in nonequilibrium systems should not be used to infer barrier heights.
\end{abstract}

\maketitle

\section{Introduction}

Activated barrier crossing is a ubiquitous phenomenon in science \cite{hanggi1990reaction, lyons2024quantifying}, such as in reaction kinetics \cite{tao2014thermally}, phase nucleation \cite{karthika2016review}, solid state defect dynamics \cite{taylor1992thermally}, fracture mechanics \cite{selinger1991statistical,zhu2006atomistic,slootman2020quantifying}, intracellular transport \cite{bressloff2013stochastic} and even earthquakes \cite{rundle2003statistical}.
Those driven by thermal fluctuations are invariably described in the framework of thermodynamics and therefore with a probability that depends on a Boltzmann weight of the relevant free-energy barrier height.
This thermodynamic picture is predicated in the context of a long-time averaging and an equilibrium ensemble picture.
Perhaps the first approach to compute a reaction rate goes back to Arrhenius with the equation \cite{hanggi1990reaction}
\begin{align} \label{eq:VHAL}
    k=k_0 \exp\left(-\frac{E_\mathrm{a}}{k_\mathrm{B}T}\right)\,,
\end{align}
where $k_0$ is a prefactor and $E_\mathrm{a}$ is the threshold energy (barrier height) for activation, $k_\mathrm{B}$ is Boltzmann's constant, and $T$ is the temperature. 
Much work has been done over the years \cite{hanggi1990reaction}, in particular with respect to determining  better estimates for the prefactor $k_0$. Even when an explicit emphasis is put on the transition path \cite{dellago1998transition, bolhuis2002transition, best2005reaction}, the rate constants are computed by means of equilibrium physics. 

In complex systems, such as condensed matter systems, the energy landscape can be complex and rugged, featuring numerous saddle points, barriers of different heights, and multiple paths connecting stable states \cite{cammarota2015spontaneous, fitzgerald2023stochastic}. 
However, it is possible to separate fast (irrelevant)  degrees of freedom from slow (relevant) ones \cite{zwanzig1961memory, hijon2010mori, te2019mori}. 
Thus, transitions in high dimensional configuration spaces \cite{satija2020broad} can often be described in terms of a simplified system with only a few degrees of freedom while the remaining `fast' degrees of freedom effectively become a stochastic fluctuating term in the resulting equations of motion.

Here, we concentrate on a minimal model for the class of systems that exhibit competing transition paths in state space, that is, a system with only two, relevant, slow degrees of freedom and also two possible transition pathways toward the global minimum.
The potential energy landscape considered here is based on that developed in Ref.~\cite{fitzgerald2023stochastic} and is illustrated in Fig.~\ref{fig:contour_plot}.
We initiate the system in one of the two local minima and study the route taken and length of time to cross over to the other (global) minimum in the energy landscape.
Systems of this type are worth considering as generic models for any stochastic process with a competition between two possible pathways.

Examples of systems with competing pathways include: In the nucleation of one phase of matter from another, two (or more) competing pathways have been observed. For example, in Ref.~\cite{yin2021transition} the various transition pathways connecting crystalline and quasicrystalline phases are discussed, while Ref.~\cite{salazar2024competing} examines the competing nucleation pathways for ZnO nanocrystal formation and the dependence on the degree of supercooling.
Somewhat related, there can sometimes be competing pathways for molecular self-assembly on surfaces.
For example, recent work examining the formation of islands of C$_{60}$ molecules on the surface of CaF$_2$ shows a competition between the pathways for single-layer and double-layer island formation \cite{holtkemeier2024nonequilibrium}.
In chemical reactions and molecular assembly, competing pathways can also arise, such as in the self-assembly of two single-stranded DNA fragments into a ring-like structure \cite{appeldorn2022employing}.

For such systems, Eq.~\eqref{eq:VHAL} suggests that the path with the lowest energy barrier $E_\mathrm{a}$ is the most probable, and thus dominates the dynamics.
Similarly, more modern reaction rate theory (RRT) \cite{hanggi1990reaction} predictions for the rate involve the exponential factor $\exp(-E_{\rm a}/k_{\rm B}T)$, albeit with more accurate estimates for the prefactor $k_0$.
For example, the Kramers--Eyring approximation for $k_0$ requires the eigenvalues of the Hessian of the underlying potential evaluated at the start-point minimum and at the barrier saddle-point \cite{bouchet2016generalisation, berezhkovskii1990solvent}.
This approximation can be very accurate, particularly at sufficiently low temperatures.
However, given that such estimates for the rate $k$ do not explicitly take into account the distance between the minima and the barrier(s), one must already conclude that considerations based solely on barrier heights and other properties local to the barrier cannot be the whole story.
Such observations motivate our investigation here of the simple model with two competing pathways.
Previous work by some of us \cite{fitzgerald2023stochastic} showed that there is a strong dependence on the time-frame over which the system is sampled, showing that on short timescales, the probability flux over the higher barrier can completely dominate the dynamics of the system.
Consequently, the observed transition probabilities significantly deviate from the predictions of RRT, {which does} not take into account situations when a finite time is available for the transition to occur, i.e.\ when the system is not in equilibrium.
RRT is applicable primarily in the long-time limit.
{Recall also that in equilibrium, microscopic reversibility dictates that transition-path times be equal for the forward and backward reactions.
In contrast, out-of-equilibrium systems break microscopic reversibility and this can be measured via the transition-path times \cite{gladrow2019experimental}.}

The conclusions of Ref.~\cite{fitzgerald2023stochastic}{, where the focus is on determining the influence of the time available on the transition rate and the route taken,} are based on the use of path integrals and via solving the Fokker--Planck equation for the time evolution of the probability density.
Here, we use Brownian dynamics computer simulations and time-dependent solutions {of} the Fokker--Planck equation, in order to explore nonequilibrium dynamical aspects of this system that are not addressed in Ref.~\cite{fitzgerald2023stochastic}.
In particular, we determine the statistics of barrier crossing locations and first-passage time distributions, amongst others.
We find that the shorter path over the higher barrier is always more likely than one would expect based on equilibrium concepts, such as using Eq.~\eqref{eq:VHAL}.
Thus, we show that these time-dependent effects in fact influence the total transition probability.
In other words, we show that one must view the transition from one state to another as a nonequilibrium process and that estimates for transition probabilities based solely on knowledge of barrier heights can be misleading, particularly for systems like that considered here, with a choice of more than one path to take.

This paper is structured as follows: In Sec.~\ref{sec:PEL} we describe and illustrate the potential energy landscape studied in this paper. In Sec.~\ref{sec:3} we briefly discuss the stochastic dynamics of our system, the Fokker--Plank equation and recall some equilibrium properties. Then, in  Sec.~\ref{sec:4} we present our results for the spatial probability distribution for the location in state space where the system crosses the barriers and in Sec.~\ref{sec:5} we present results for the length of time taken to arrive at the destination. In Sec.~\ref{sec:6} we present results from solving the Fokker--Plank equation. Finally, in Sec.~\ref{sec:7}, we make a few concluding remarks.

\section{Potential Energy Landscape}\label{sec:PEL}

\begin{figure} 
\centering
\includegraphics[width=0.5\textwidth]{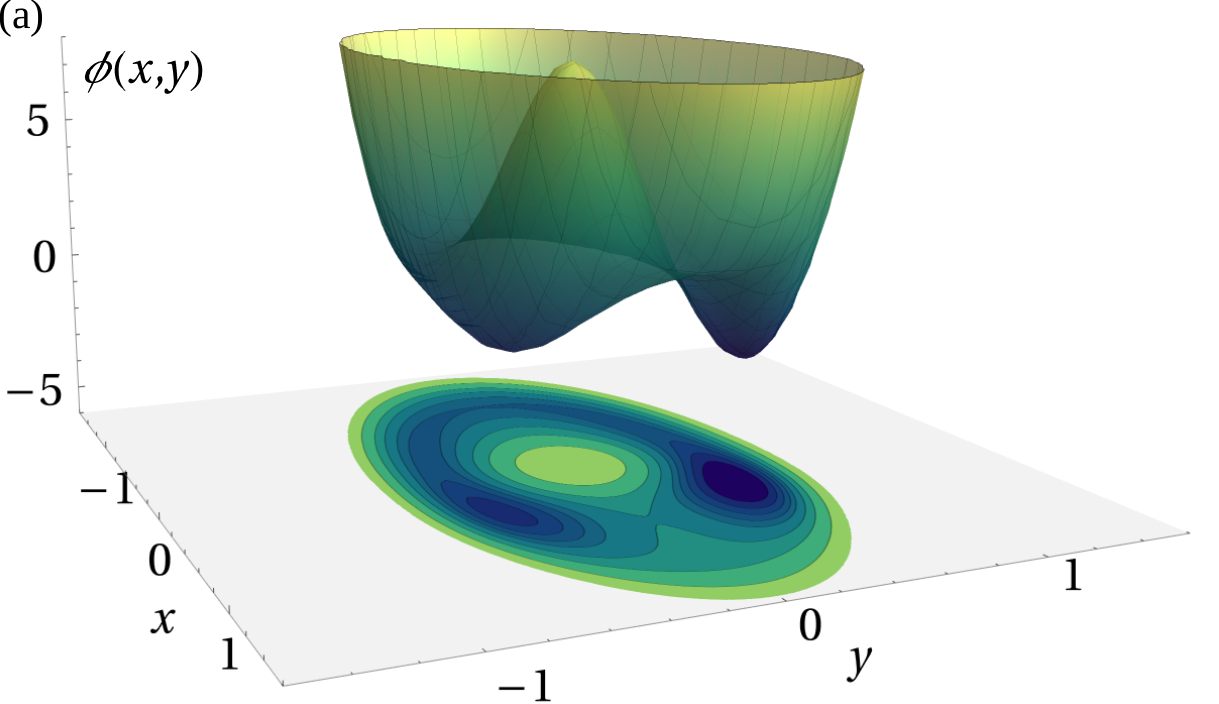}
\includegraphics[width=0.46\textwidth]{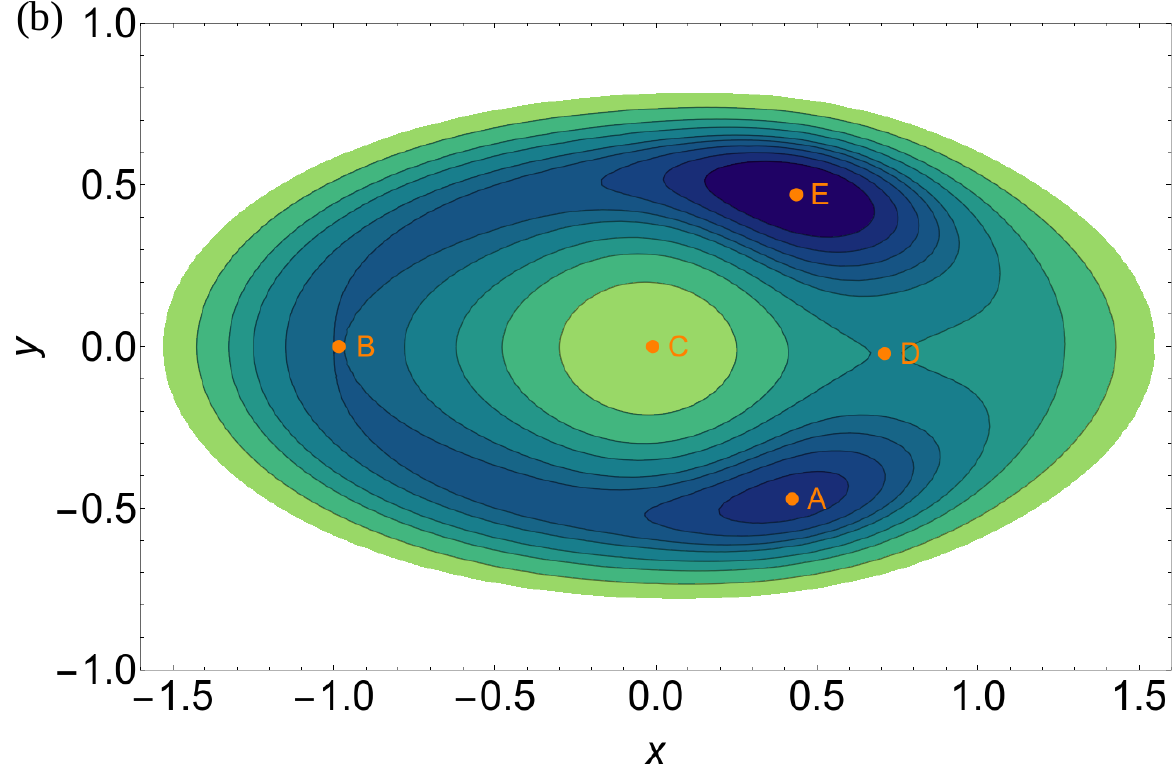}
\caption{(a) Plot of the surface described by the potential $\phi(x,y)$ in Eq.~\eqref{eq:potential}, together with its projection on the $xy$-plane. (b) Contour plot of the potential $\phi(x,y)$, with the stationary points indicated.}
\label{fig:contour_plot}
\end{figure}

\begin{figure*}
    \centering
    \includegraphics[width=.49\textwidth]{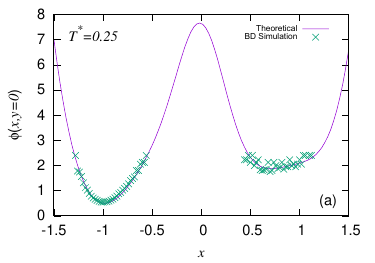}
    \includegraphics[width=.49\textwidth]{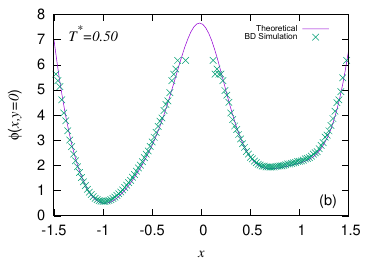}
    \includegraphics[width=.49\textwidth]{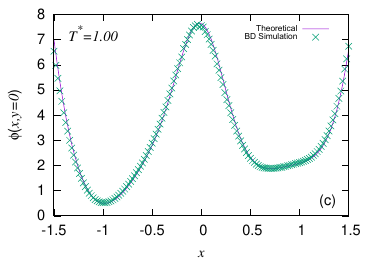}
    \includegraphics[width=.49\textwidth]{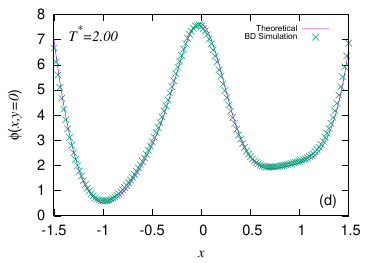}
    \caption{Plots of $\phi(x,y=0)$, i.e.\ cuts through the potential in Eq.~\eqref{eq:potential} for various temperatures (as indicated), displayed in each case as the solid line. We also present the results from Brownian dynamics (BD) simulations, sampling the density distribution $\rho(x,y=0)$, which are displayed (using $\times$ symbols) by plotting the quantity $-k_BT \ln \rho(x,y=0)+$constant [c.f.\ Eq.~\eqref{eq:ln_rho}]. Each is the result from a single simulation of total time 2000$\tau_B$.}
    \label{fig:Equilibrium_plot}
\end{figure*}

We study a two-dimensional system, with coordinates $\boldsymbol{x}=(x,y)$, exploring the potential energy landscape 
\begin{equation}\label{eq:potential}
   \phi(x,y)=\phi_{sM}(x,y)-{b}\,k_\mathrm{B}T_\mathrm{ref}\frac{x}{\ell}+\phi_G(x,y),
\end{equation}
where $T_\mathrm{ref}$ is a reference temperature, $\beta_\mathrm{ref}\equiv (k_\mathrm{B} T_\mathrm{ref})^{-1}$, and $\ell$ is the length-scale in our model. 
Figure~\ref{fig:contour_plot}(a) shows the surface defined by $\phi(x,y)$ in Eq.~\eqref{eq:potential}, while Fig.~\ref{fig:contour_plot}(b) shows the contour plot of the projection of  $\phi$ on the $xy$-plane, with the stationary points where $\nabla\phi=0$ indicated.
Henceforth, we define a dimensionless temperature $T^*=T/T_\mathrm{ref}$ and set the unit of length $\ell=1$.
The first term in Eq.~\eqref{eq:potential} is a stretched Mexican-hat type potential
\begin{equation}
    \beta_\mathrm{ref}\phi_{sM}(x,y)=4 \left(x^{2}+4 y^{2}-1\right)^{2}\,{.}
\end{equation}
{For $b>0$, the second term in Eq.~\eqref{eq:potential} leads to a small constant force parallel to the $x$-axis.
Unless otherwise stated, we set here the value of this parameter to be $b=0.5$. However, in Sec.~\ref{sec:5} we present results for a range of different values of $b$. The final term in Eq.~\eqref{eq:potential} is}
\begin{align} \label{eq:pot}
 \beta_\mathrm{ref} \phi_G(x,y) = & -2 {\mathrm e}^{-4 \left(x -\frac{1}{2}\right)^{2}-4 \left(y -\frac{1}{2}\right)^{2}} \nonumber \\
&-{\mathrm e}^{-4 \left(x -\frac{1}{2}\right)^{2}-4 \left(y +\frac{1}{2}\right)^{2}} \nonumber \\
 &+3{\mathrm e}^{-4 \left(x -1\right)^{2}-4 y^{2}}+4 {\mathrm e}^{-10 \left(x^{2}+y^{2}\right)}{,}
\end{align}
{which} consists of four Gaussian contributions to generate local maxima and minima in the potential at various points. The final term in $\phi_G$ was not included in the work in Ref.~\cite{fitzgerald2023stochastic}.
Here, we add this term to make the local maximum at the origin even higher and thereby decrease the probability of the system to pass near this point $C$ [see Fig.~\ref{fig:contour_plot}(b)] and to instead more often take either of the two paths near the saddle points $B$ and $D$.

The five stationary points of $\phi(x,y)$ have coordinates:
\begin{align}
\label{eq:points}
    A&=(x_A, y_A)=(0.413, -0.471)\,,\nonumber \\
    B&=(x_B, y_B)=(-0.984, -0.000)\,,\nonumber \\
    C&=(x_C, y_C)=(-0.017, 0.003)\,,\nonumber \\
    D&=(x_D, y_D)=(0.710, -0.022)\,,\nonumber \\
    E&=(x_E, y_E)=(0.436, 0.460).
\end{align}
Point $A$ is a local minimum, while $E$ is the global minimum. Point $C$ is a local maximum and points $B$ and $D$ are saddle points.

\begin{figure*} 
    \centering
    \includegraphics[width=.49\textwidth]{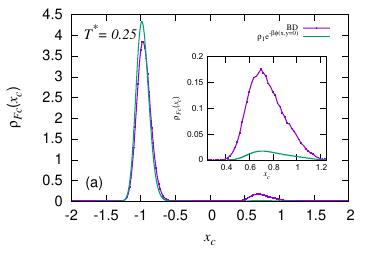}
    \includegraphics[width=.49\textwidth]{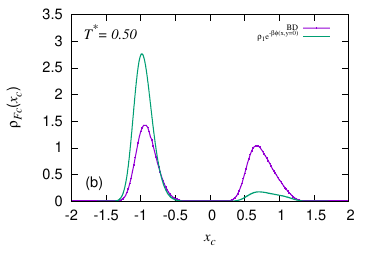}
    \includegraphics[width=.49\textwidth]{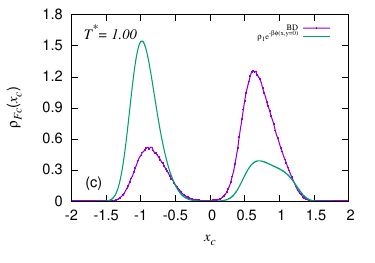}
    \includegraphics[width=.49\textwidth]{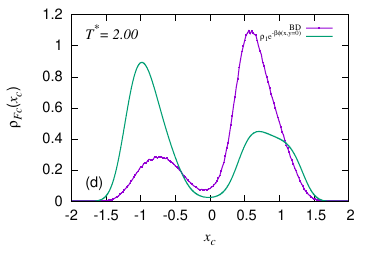}
    \caption{Probability density $\rho_{FC}(x_c)$ over the position $x_c$ where the system finally crosses the $x$-axis, having been initiated at point $A$ in Fig.~\ref{fig:contour_plot}(b), calculated using BD simulations. Each run is terminated when it reaches the vicinity of the global minimum, point $E$. Panels (a)--(d) present results for the four different temperatures indicated and compare with the equilibrium density result in Eq.~\eqref{eq:epd}, as the solid green line in each case.}
    \label{fig:x_c_histogram}
\end{figure*}

The scenario that we consider here is to initiate the system at point $A$ and then consider the route and time taken for it to evolve to the global minimum at $E$.
We consider the system's coordinates $(x,y)$  to undergo Brownian motion.
Two typical pathways from $A$ to $E$ exist.
The first pathway involves passing in the vicinity of saddle point $D$; the second pathway goes through the vicinity of point $B$, the other saddle point.
Passing near $B$ requires the system to traverse a greater distance, resulting in a longer reaction pathway.

The barrier height to be surmounted going from the initial point $A$ to the saddle at point $D$ is $\Delta \phi_{AD}\equiv\phi(x_D,y_D)-\phi(x_A,y_A)=2.7k_\mathrm{B}T_\mathrm{ref}$. In contrast, the barrier height for going up to the saddle at point $B$ from initial point $A$ is $\Delta \phi_{AB}\equiv\phi(x_B,y_B)-\phi(x_A,y_A)=1.3k_\mathrm{B}T_\mathrm{ref}$.
Thus, the system is presented with the choice to (i) go over near the higher saddle at point $D$, with a relatively short distance to travel in state space, or (ii) go over near the lower saddle point $B$.
However, this route involves a longer journey.
On the basis of Eq.~\eqref{eq:VHAL} (with $E_\mathrm{a}=\Delta\phi_{AD}$ or $E_\mathrm{a}=\Delta\phi_{AB}$), and the assumption that the prefactor is similar for both routes, one would predict that the path going via point $B$ is significantly more likely than the path via the saddle at $D$.
It turns out, as we show below, this is not what in reality occurs, particularly when the time available for the transition to occur is a factor in the problem.

\section{Stochastic Dynamics}
\label{sec:3}

We assume that the time evolution of the system is given by the following overdamped stochastic dynamical equation \cite{Risken}:
\begin{equation} \label{eq:OvD}
    \gamma \Dot{\boldsymbol{x}} = -\nabla \phi(\boldsymbol{x}) + \boldsymbol{\xi}(t),
\end{equation} 
where $\gamma$ is the friction coefficient and the force $\boldsymbol{\xi}$ is a Gaussian white noise.
In conjunction with the nondimensionalisation introduced above in our discussion of the potential in Eq.~\eqref{eq:potential}, where we choose $\ell$ and $k_BT_{\rm ref}$ as the length and energy scales in which to work, we now introduce the natural (Brownian) timescale for the system,
{\begin{equation}\label{eq:tau_B}
\tau_B=\beta_{\rm ref}\gamma\ell^2.
\end{equation}
}
Thus, the random components of $\boldsymbol{\xi}(t)$ obey
\begin{align} 
        \langle \xi_i (t) \rangle &=0 \,, \label{eq:EA1}\\
 \langle \xi_i (t) \xi_j(t') \rangle &=2\gamma k_\mathrm{B} T \delta_{ij}\delta(t-t')\,, \label{eq:EA2}
 \end{align}
where $\delta(t-t')$ is a Dirac delta distribution, $\delta_{ij}$ is the Kronecker delta, with $i,j =\{x,y\}$, and $\langle \cdot \rangle$ denotes the ensemble average.
{Note that Eqs.~\eqref{eq:OvD}--\eqref{eq:EA2} are predicated on the hypothesis of Markovian process, i.e., that there is no memory in the system.
This is not always a trivial issue.
A test for Markovianity along a reaction coordinate has been devised in Ref.~\cite{berezhkovskii2018single}.}
We solve Eq.~\eqref{eq:OvD} numerically using the standard Euler--Maruyama finite difference method \cite{2R13}, with a time-step $\Delta t=10^{-3}$.

One can also determine properties of the dynamics of the system by solving the Fokker--Planck (Smoluchowski) equation for the time evolution of $\rho(\boldsymbol{x},t)$, the probability density for finding the system at point $\boldsymbol{x}$ at time $t$. This partial differential equation is \cite{Risken}
\begin{align} \label{eq:Smolu}
   {\frac{1}{{\cal D}}} \frac{\partial \rho}{\partial t}=\nabla^2\rho+\nabla\cdot\left[\rho\nabla({\beta}\phi)\right],
\end{align}
{where ${\cal D}=k_BT/\gamma$ is the diffusion coefficient.
Thus, at our reference temperature $T_{\rm ref}$ we have ${\cal D}=1$.
Equation~\eqref{eq:Smolu}} must be solved together with an initial condition $\rho(\boldsymbol{x},t=0)$, which is specified below.
The equilibrium (in the limit $t\to\infty$) probability density is the Boltzmann distribution
\begin{align} \label{eq:rho_eq}
    \rho(x,y)=\rho_0e^{-\beta \phi(x,y)},
\end{align}
where $\rho_0^{-1}=\int_{-\infty}^{\infty}\int_{-\infty}^{\infty}e^{-\beta \phi(x,y)}dxdy$ is the normalisation constant.
{Note that in Eq.~\eqref{eq:rho_eq} and henceforth, whenever we discuss the $t\to\infty$ equilibrium limit, we omit the time dependence of the density $\rho$.}
We use this exact result to check the performance of our numerical method for solving the stochastic dynamics and to determine the temperature range in which we can work with the computational resources available. Rearranging Eq.~\eqref{eq:rho_eq}, we obtain the following expression for the potential $\phi$ in terms of the density:
\begin{align}
\label{eq:ln_rho}
     \phi(x,y)=-k_BT \ln \rho(x,y)+k_BT \ln \rho_0\,,
\end{align}
where the last term on the right-hand side is a constant.
Therefore, by sampling the dynamics from Eq.~\eqref{eq:OvD} to obtain $\rho(x,y)$ and then by plotting $-k_BT \ln \rho+$constant and comparing with $\phi$, we have a good check for the accuracy of our simulations, and also a method to identify roughly what sort of timescales are needed to obtain the expected distribution (relevant to the results presented later).

Figure \ref{fig:Equilibrium_plot} shows the cross-section of the potential $\phi(x,y=0)$. 
We also show Brownian dynamics (BD) computer simulations of  Eq.~\eqref{eq:OvD} at the temperatures $T^*=0.25$, $0.5$, $1$, and $2$ to sample $\phi$ from Eq.~\eqref{eq:ln_rho}.
Note that the two saddle points $B$ and $D$ lie almost on the line $y=0$, as well as the local maximum at point $C$. The occupation density $\rho(x,y=0)$ is built from a single, long run at each temperature value, lasting for a total time $2000\tau_B$.
The data are sampled at high frequency during each simulation.
We see that, in agreement with the typical expectation, during the simulation, the system crosses the saddle points (which are the minima in Fig.~\ref{fig:Equilibrium_plot}) repeatedly and, at large $T$ the state space is well-sampled.
However, for the lower temperatures $T^*=0.25$ and 0.5, the system does not reach the local maximum at $x\approx0$, so this part of configuration space is not well-sampled at these temperatures. Longer simulation runs are required to observe the system at this point for temperatures $T^*\lesssim 0.5$.
Note that RRT approximations for the transition rates, such as the Kramers--Eyring result \cite{bouchet2016generalisation, berezhkovskii1990solvent}, assume that barrier crossing occurs close to the saddle points. That only the barrier regions are well-sampled in the low temperature BD simulations in Fig.~\ref{fig:Equilibrium_plot} is an indicator that barrier crossing does indeed occur close to the saddle points (local minima in Fig.~\ref{fig:Equilibrium_plot}) for these temperatures.
However, at higher temperatures we see that even the local maximum at $x=0$ is well-sampled, indicating that for these higher temperatures, barrier crossing away from the saddle points occurs.
This suggests that any approximation for transition rates that assumes barrier crossing  solely occurs at the saddles will start to fail for these higher temperatures, i.e.\ at $T^*>0.25$ for the system here.
This figure also shows that (as implemented) our BD simulation approach cannot be used at lower temperatures.

\section{Crossing statistics}
\label{sec:4}

Having confirmed that our methodology is correctly sampling the equilibrium probability density for $T^*\geq0.25$ and having gained some initial insight into some of the relevant timescales governing the system, we now discuss the nonequilibrium probability densities.
The situation we consider is relevant to all process where the system is stopped as soon as the barrier-crossing process has occurred. 
For example, in the case of chemical reactions, this is the situation where the product molecules are removed, as soon as they are produced. 

As alluded to in the introduction, we initiate the system at the local minimum point $A$ -- see Fig.~\ref{fig:contour_plot}(b) and Eq.~\eqref{eq:points} -- and stop the system when it reaches the vicinity of the global minimum, point $E$, that is, when $|\boldsymbol{x}-\boldsymbol{x}_E|<0.1$.
For this process of crossing barriers from $A$ to $E$, the point along the $x$-axis where the system crosses and also the time taken to reach the destination are two interesting and informative quantities.
In this section, we discuss the statistics of the crossing point, $x_c$. We discuss the travel times in Sec.~\ref{sec:5}.

{\subsection{Crossing location}}

Our definition of the $x$-axis crossing point $x_c$ needs some clarification.
This is because it is entirely possible for the system to cross back and forth across the line $y=0$ multiple times, before it finally heads down the potential well towards the destination global minimum point $E$.
Thus, we define the $x$-axis crossing point $x_c$ as the $x$-coordinate value at the latest time when $y=0$, before reaching the destination.
Or, to be more precise, given that we solve our dynamics in Eq.~\eqref{eq:OvD} using finite differences in time, $x_c$ is defined as the $x$-value at the last time when the $y$-coordinate changes sign.

\begin{figure}[t!]
    \centering
    \includegraphics[width=.49\textwidth]{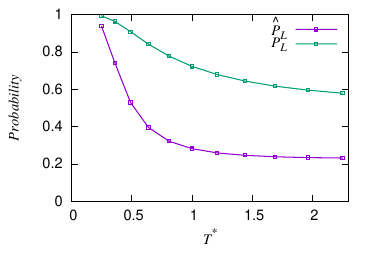}
    \caption{The probability $\hat{P}_L$, defined in Eq.~\eqref{eq:hatP}, plotted as a function of temperature, calculated using BD simulations. $\hat{P}_L$ is the probability for the system initiated at point $A$ [see Fig~\ref{fig:contour_plot}(b)] to move to the global minimum at $E$, while crossing the $x$-axis to the left of the origin. We also display $P_L$, the estimate based on the equilibrium density distribution, defined in Eq.~\eqref{eq:P_eq}. Of course, the corresponding right-crossing probabilities are $\hat{P}_R=1-\hat{P}_L$ and $P_R=1-P_L$.}
    \label{fig:pf}
\end{figure}

In Fig.~\ref{fig:x_c_histogram} we plot the density distribution $\rho_{Fc}(x_c)$, which is the histogram for the final crossing point $x_c$.
We use 200 equally spaced bins over the interval $x_c\in[-2,2]$ to sample $\rho_{Fc}(x_c)$. The distribution $\rho_{Fc}(x_c)$ is normalised so that $\int_{-\infty}^\infty\rho_{Fc}(x_c)dx_c=1$.

\begin{figure*} 
    \centering
    \includegraphics[width=.49\textwidth]{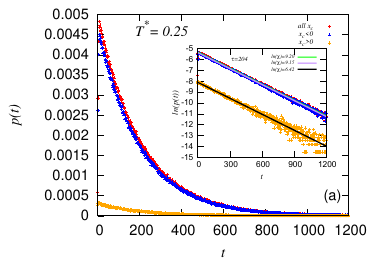}
    \includegraphics[width=.49\textwidth]{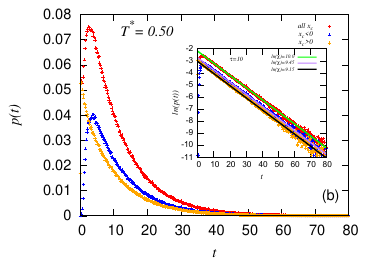}
    \includegraphics[width=.49\textwidth]{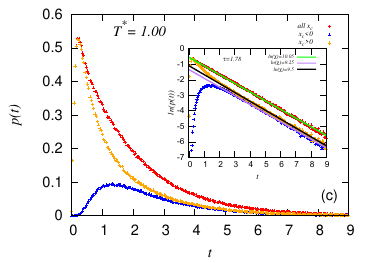}
    \includegraphics[width=.49\textwidth]{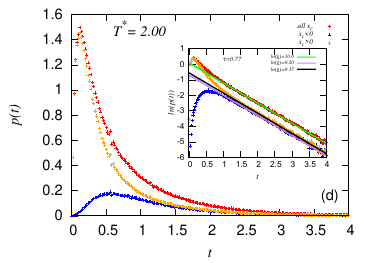}
    \caption{Arrival time distributions (histograms) for the four different temperatures indicated. In each case, the red symbols are the distribution irrespective of the value of $x_c$, the point on the $x$-axis where the system crossed, while the blue and orange distributions are for $x_c<0$ and $x_c>0$, respectively. In the insets, we plot $\ln p(t)$ versus $t$ using the same data as displayed in the main plots, in order to display as a straight line the exponential tails. We also display straight line fits to the tails, i.e.\ fits to Eq.~\eqref{eq:exp_distrib}.}
    \label{fig:seg}
\end{figure*}

In view of Eq.~\eqref{eq:rho_eq}, it is perhaps natural to guess that a good estimate for $\rho_{Fc}(x_c)$ is to assume $\rho_{Fc}(x_c)\propto \rho(x_c,y=0)$.
Thus, in Fig.~\ref{fig:x_c_histogram} we also plot the distribution
\begin{align} \label{eq:epd}
    \rho_{eq}(x)= \rho_1e^{-\beta \phi(x,y=0)},
\end{align}
with $\rho_1^{-1}=\int_{-\infty}^{\infty}e^{-\beta \phi(x,y=0)}dx$.
It is important to mention that the normalisation constants $\rho_0\neq \rho_1$.
In Fig.~\ref{fig:x_c_histogram} we compare $\rho_{Fc}(x_c)$ with $\rho_{eq}(x)$, showing the results for the temperatures $T^{*}=0.25$, 0.5, 1 and 2.

Figure~\ref{fig:x_c_histogram}(a) shows $\rho_{Fc}(x_c)$ for the lowest temperature considered, $T^*=0.25$.
The largest peak in $\rho_{Fc}(x_c)$ is centered near $x_c\approx-1$, indicating the majority cross near the lowest energy saddle point $B$, which is what one would expect based on Eq.~\eqref{eq:VHAL}.
However, there is a second noticeable peak centered around $x_c\approx0.7$, corresponding to the system crossing near the higher saddle point $D$.
A magnification of this peak is shown in the inset.
What is particularly striking is that here the density $\rho_{Fc}(x_c)\gg\rho_{eq}(x_c)$, indicating that the probability of the transition occurring via the higher barrier is much higher than one would estimate based on assuming Eq.~\eqref{eq:epd}.

Moving on to the higher temperature results in Fig.~\ref{fig:x_c_histogram}(b)--(d), we see that as the temperature is increased, for $x_c>0$ the density $\rho_{Fc}(x_c)$ is always greater than $\rho_{eq}(x_c)$.
In other words, going from $A$ to $E$ over the higher barrier near saddle $D$ is an event with much higher probability than one would expect from comparing the Boltzmann factors from Eq.~\eqref{eq:rho_eq} at the two saddle points.
At $T^*=0.5$ [Fig.~\ref{fig:x_c_histogram}(b)], the area under the two peaks is roughly equal, indicating the probability for going from $A$ to $E$ over the barrier near $B$ is roughly the same as the probability for going over near $D$.
This is not what one would expect based on Eq.~\eqref{eq:VHAL}.
Moreover, for the even higher temperatures in Fig.~\ref{fig:x_c_histogram}(c) and (d), the area under the right-hand peak centered at $x_c\approx0.7$ is greater than the area under the peak at $x_c\approx-1$.

Another trend that can be observed from Fig.~\ref{fig:x_c_histogram} is that as the temperature of the system is increased, as to be expected, the width of the peaks centered around the two saddle points at $x_c\approx-1$ and $x_c\approx0.7$ become broader and at the higher temperatures the probability of being at the local maximum $C$ with $x_c\approx0$ becomes sizable.

\subsection{Left- and right-crossing probabilities}

In view of the above observations, it is informative to calculate the total left- and right-crossing probabilities $\hat{P}_L$ and $\hat{P}_R$ for the system to cross the $x$-axis on the negative half, and on the positive half, respectively
\begin{align} \label{eq:hatP}
\begin{split}
    \hat{P}_L=& \int_{-\infty}^0\rho_{Fc}(x_c)dx_c \\
    \hat{P}_R=& \int_0^{\infty}\rho_{Fc}(x_c)dx_c \,,
\end{split}
\end{align} 
where $\rho_{Fc}(x_c)$ is the crossing point distribution (histogram) discussed above, with examples presented in Fig.~\ref{fig:x_c_histogram}. Of course, $\hat{P}_L+\hat{P}_R=1$.
Corresponding equilibrium statistical mechanics estimates for these probabilities are
\begin{align} \label{eq:P_eq}
\begin{split}
    P_L=& \int_{-\infty}^0\rho_{eq}(x)dx  \\
    P_R=& \int_0^{\infty}\rho_{eq}(x)dx \,,
\end{split}
\end{align} 
where $\rho_{eq}(x)$ is given in Eq.~\eqref{eq:rho_eq}.
In Fig.~\ref{fig:pf} we present results for $\hat{P}_L$ and $P_L$, for a range of temperatures.
{These results for $\hat{P}_L$ and $\hat{P}_R$ are obtained from sampling over $2\times10^6$ independent BD simulations at each temperature value.
Note that the statistical errors are smaller than the symbol size.}
Figure \ref{fig:pf} shows explicitly what one can infer from Fig.~\ref{fig:x_c_histogram}, namely, that the probability to go from $A$ to $E$ via the higher barrier $D$ to the right is always higher than one would expect based on the equilibrium estimate, Eq.~\eqref{eq:P_eq}.
This higher probability is due to the shorter distance in configuration space of the pathway through $D$. 
{Therefore, a RRT that takes into account the shape of $\phi$ to determine the prefactors $k_0$ in Eq.~\eqref{eq:VHAL} for the two different paths will give better account than Eq.~\eqref{eq:P_eq} for the relative probabilities to take the two paths.}

At high temperatures, it is perhaps not surprising that the system should take the shorter route, but what is interesting is that even at very low temperatures, such at $T^*=0.25$, there is a sizable probability flux over the higher barrier, with a much higher probability than equilibrium arguments of the kind leading to Eq.~\eqref{eq:P_eq} would predict.
While the path over the higher barrier is more likely if there is only a short time available for the transition to occur, what the present results show is that even when no time-limit is imposed, there is still a nonequilibrium bias on the overall transition probability, stemming from how we initiate the system.

\section{Travel Time Statistics}
\label{sec:5}

\begin{figure*}[t!] 
\centering
    \includegraphics[width=0.32\linewidth]{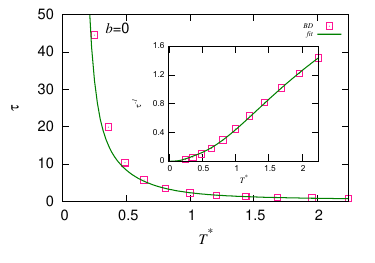} 
    \includegraphics[width=0.32\linewidth]{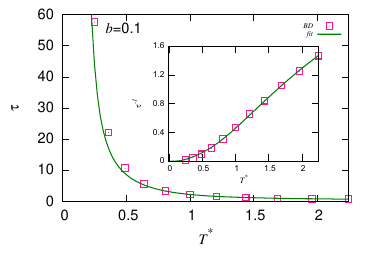}
    \includegraphics[width=0.32\linewidth]{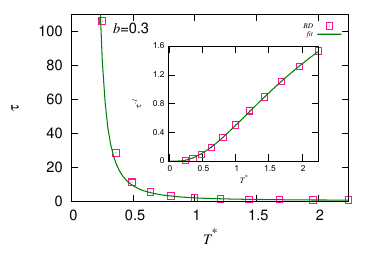}
    \includegraphics[width=0.32\linewidth]{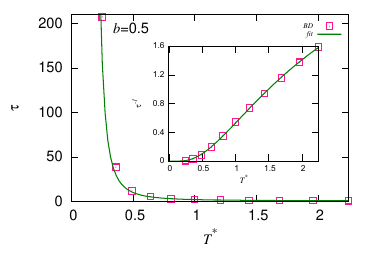}
    \includegraphics[width=0.32\linewidth]{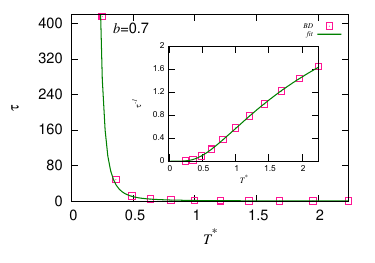}
    \includegraphics[width=0.32\linewidth]{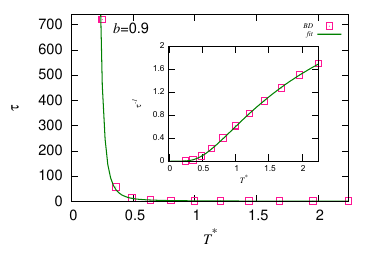}
\caption{Plot of the decay time $\tau$ in Eq.~\eqref{eq:exp_distrib} as a function of temperature{, for various values of the external potential parameter $b$ (as indicated on each plot)}, obtained from BD (symbols) by fitting the tails of the distributions, like in the insets of Fig.~\ref{fig:seg}. {The bottom left plot for $b=0.5$ is the case corresponding to Fig.~\ref{fig:seg} and all previous results. In each case, the inset} shows the reciprocal $\tau^{-1}$ versus temperature. The solid {lines in all plots} is the fit to the BD data using Eq.~\eqref{eq:15}.}
\label{fig:tau_tau-1}
\end{figure*}

Transition path times have attracted considerable interest in recent years \cite{hummer2004transition,berezhkovskii2005one,Dudko2006,sega2007,zhang2007transition,chung2009experimental,chaudhury2010harmonic,orland2011generating,kim2015mean,makarov2015shapes,Truex2015,daldrop2016transition,pollak2016transition,Neupane2016,caraglio2018influence,Neupane2018} because these times give information about the structural changes in, e.g., protein folding transitions \cite{cossio2018transition}. However, the vast majority has focused on one-dimensional energy landscapes. 

In our two-dimensional system, after having discussed the statistics of the crossing point $x_c$, we now move on to discuss the corresponding distributions of arrival times, i.e.\ as mentioned above, the time taken till the system coordinates satisfy the condition $|\boldsymbol{x}-\boldsymbol{x}_E|<0.1$, having been initiated at point $A$ at time $t=0$.
The arrival times discussed below are related to the first-passage time problem \cite{redner2023first} and the general problem of transition-path theory {\cite{vanden2010transition,vanden2006transition,vanden2006towards}.}

Figure~\ref{fig:seg} shows the probability density distribution of arrival times $p(t)$ for the temperatures $T^{*}=0.25$, $0.5$, 1, and 2, and also the respective densities split according to whether the system traveled to the destination crossing the $x$-axis to the left or to the right of the origin.
These results are  averages over $2\times10^6$ independent BD simulations for each temperature.
The number of bins used for the histograms are $10^3$, $10^4$, $5\times10^4$, and $10^5$ for each temperature, respectively. 
The first thing that can be seen from the plots in Fig.~\ref{fig:seg}, just from comparing the ranges of the four different horizontal $t$-axes, is that the typical time to travel from $A$ to $E$ decreases with increasing temperature.
This is not surprising.
However, what is perhaps surprising is how relatively simple and structureless $p(t)$ seems to be.
Given the complex spatial behaviour discussed in the previous section, one might expect $p(t)$ to perhaps exhibit features corresponding to the two different paths available for the system to cross from $A$ to $E$.
{Interestingly, the plots in Fig.~\ref{fig:seg} also show that at larger times, the two partial distributions (for $x_c>0$ and $x_c<0$, respectively) decay exponentially, with the same decay rate $\tau$ as the total distribution (for all $x_c$).}

{To see this decay more clearly, in} the Fig.~\ref{fig:seg} insets, we plot the logarithm $\ln(p(t))$ versus time, for the same data as in the main plots.
The fact that this forms a straight line over a large part of the range (specifically in the intermediate- and long-time $t$ regime), shows that except for very short times, $p(t)$ follows the exponential distribution
\begin{align}
    p(t)=\chi e^{-t/\tau},
    \label{eq:exp_distrib}
\end{align}  
where the prefactor $\chi\approx1/\tau$. The prefactor is not exactly equal to $1/\tau$, due to the deviations from Eq.~\eqref{eq:exp_distrib} at short times.

It can be understood why the probability density $p(t)$ follows an exponential distribution in the long-time limit by considering simple rate equations. In the long-time limit, the rates for crossing each of the barriers, $k_{AB}$ and $k_{AD}$, can be assumed to be constants and so the probability density for the transition from the initial metastable state can be obtained from the equation $\partial p(t)/\partial t=-(k_{AB}+k_{AD})p(t)$, so that the solution is an exponential with a single time $1/(k_{AB}+k_{AD})$. 
Another way to understand why Eq.~\eqref{eq:exp_distrib} has an exponential form comes from identifying our dynamics as one with a form of `stochastic resetting', where such exponentials can arise \cite{2R16}. The sequence of simulations performed to sample the distribution $p(t)$ [e.g.\ in Fig.~\ref{fig:seg}] may equivalently be viewed as a single run, with a stochastic resetting occurring when each run reaches the destination and then the next run is started. Thus, it is perhaps not surprising that for larger times $p(t)$ has an exponential form.

\begin{table}{
    \centering
    \begin{tabular}{ccccc}
      $b$   & ~ ${\beta_{\rm ref}}\Delta\phi_{AB}$~ & ~ ${\beta_{\rm ref}}\Delta\phi_{AD}$ & ~ $c_{AB}\tau_B$ & ~ $c_{AD}\tau_B$\\ 
      \hline 
       0.0  & 0.62 & 2.85 & 0.36 & 4.12\\
       0.1  & 0.75 & 2.82 & 0.46 & 3.98\\
       0.3  & 1.03 & 2.76 & 0.75 & 3.61\\
       0.5  & 1.31 & 2.70 & 1.21 & 3.01\\
       0.7  & 1.59 & 2.64 & 2.08 & 1.96\\
       0.9  & 1.87 & 2.58 & 3.88 & 0.00\\
    \end{tabular}
    \caption{The barrier heights $\Delta\phi_{AB}$ and $\Delta\phi_{AD}$ and corresponding {prefactors $c_{AB}$ and $c_{AD}$ in Eq.~\eqref{eq:15} ($\tau_B$ is the Brownian timescale in Eq.~\eqref{eq:tau_B})}, obtained from fitting the data displayed in Fig.~\ref{fig:tau_tau-1}, for varying $b$ in the external potential~\eqref{eq:potential}.}
    \label{tab:bh_ce}}
\end{table}

In Fig.~\ref{fig:seg} we also display (straight-line) fits to the tails of the data, performed using Fisher's scoring method \cite{2R18}, a variant of Newton's method.
This {helps us to see clearly the decay timescales are all the same and also allows us to} extract the value of the decay timescale $\tau$, which we plot as a function of temperature in Fig.~\ref{fig:tau_tau-1}. {We see that upon decreasing $T$, the timescale $\tau$ rapidly increases.}

\begin{figure*}[t!]
   \centering
    \includegraphics[width=.8\textwidth]{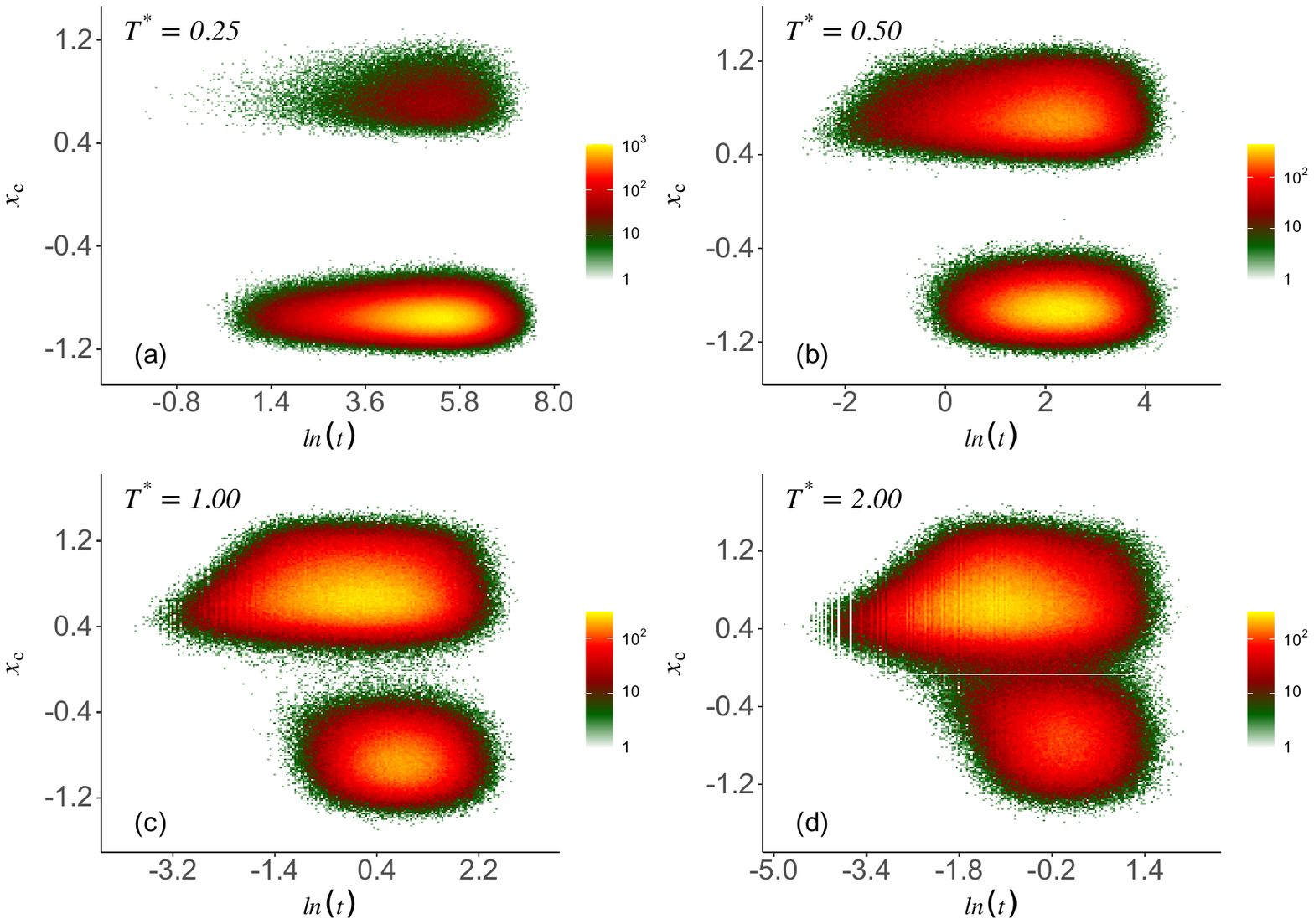}
    \caption{Histogram of the $x$-axis crossing point $x_c$ versus the logarithm of the arrival time, $\ln(t)$, for a total of $2\times10^6$ different independent runs, for four different temperatures, as indicated. Note the log-scale of the histogram colour bar.}
    \label{fig:log_t_x_cross}
\end{figure*}

Note that in Fig.~\ref{fig:tau_tau-1}, we not only plot the results for the $b=0.5$ case, but also for several other values of $b$.
Recall that the parameter $b$ determines the strength of the linear term in the external potential \eqref{eq:potential}. Until this point in the paper we have solely considered the $b=0.5$ case, but we now present results for a range of values $b$.
Varying $b$ allows to easily alter the relative heights of the potential $\phi$ at saddle points $B$ and $D$, as can be seen from the selection of barrier height values given in Table~\ref{tab:bh_ce}.
As $b$ is increased, then so is the value of $\Delta\phi_{AB}$, while the value of $\Delta\phi_{AD}$ is decreased. Note that for $b<0$, the saddle point at $B$ is replaced by a local minimum, with two new saddle points appearing either side of it, completely changing the character of the potential.
Thus, we do not consider here the case $b<0$.
This change of character from saddle to local minimum at point $B$ occurring at $b\approx0$ can also be seen from calculating the eigenvalues $\lambda_+$ and $\lambda_-$ of the Hessian of the potential $\phi(x,y)$ at this point.
At $B$, the positive eigenvalue $\lambda_+ \approx31$, hardly changing in value for $b\in[0,0.9]$, while the negative eigenvalue $\lambda_-\approx0$ for $b=0$, decreasing to $\lambda_- \approx-2$ and then $\lambda_- \approx-3.7$, for $b=0.5$ and $b=0.9$, respectively.
Also, for $b\approx 0.9$ the heights of the two barriers become similar and therefore essentially all of the flux from $A$ to $E$ goes via the shorter route near saddle $D$, so higher $b$ values are not worth considering.
At saddle point $D$ there are modest changes in the eigenvalues with varying $b$: the positive eigenvalue $\lambda_+ \approx16$ for $b=0$, changing to $\lambda_+ \approx12$ for $b=0.5$. Similarly, the negative eigenvalue $\lambda_- \approx-59$ for $b=0$ and $\lambda_- \approx-56$ for $b=0.5$.
One can use these eigenvalues together with the ones calculated at the start-point~$A$ as input to the Kramers--Eyring expression for the rates over each of the barriers \cite{bouchet2016generalisation, berezhkovskii1990solvent}. For the low temperature $T^*=0.25$ and when $b=0.5$ the ratio of the rates from Kramers--Eyring is $k_{AB}/k_{AD}\approx31$, which is roughly double the observed ratio $\approx 15$ (see Fig.~\ref{fig:pf}). However, even at this temperature, Kramers--Eyring is starting to fail because the barriers are relatively small. Thus, systematic comparison with the results reported here is not profitable, because it is not possible to distinguish whether Kramers--Eyring is failing because the barriers being considered are only modest in height or whether it is because of the path-length aspects and other issues that are the focus here.
Recall that, as remarked already in our discussion of Fig.~\ref{fig:Equilibrium_plot}, with the BD simulation methods used here, properly sampling at low temperatures (higher barriers) becomes unreliable too.

Figure~\ref{fig:tau_tau-1} also displays the best fit to the data {for the various values of $b$} using the following expression for the rate constant [c.f.\ Eq.~\eqref{eq:VHAL}]:
\begin{equation}
\frac{1}{\tau}=c_{AB}\exp\left(-\frac{\Delta\phi_{AB}}{k_\mathrm{B}T}\right)+c_{AD}\exp\left(-\frac{\Delta\phi_{AD}}{k_\mathrm{B}T}\right).
\label{eq:15}
\end{equation}
The prefactors $c_{AB}$ and $c_{AD}$ are treated as fitting parameters.
Using the least squares method, we find the values $c_{AB}=1.21\tau_\mathrm{B}^{-1}$ and $c_{AD}=3.01\tau_\mathrm{B}^{-1}${, where $\tau_B$ is the timescale in Eq.~\eqref{eq:tau_B},} give the best fit to the data {for the case where $b=0.5$. In Table~\ref{tab:bh_ce} we give the values obtained for the fitting parameters $c_{AB}$ and $c_{AD}$ for the various values of $b$ considered in Fig.~\ref{fig:tau_tau-1}.
Equation~\eqref{eq:15}} essentially assumes the total rate is the sum of two independent rates, each given by an Arrhenius-type expression. As we show below [see Fig.~\ref{fig:log_t_x_cross}], these two processes are not independent of each other, but the fact that Eq.~\eqref{eq:15} fits the data so well, suggests the coupling between the two processes is weak, which tallies with the fact that we are dealing with rare events. {Note that for $b\geq 0.9$ we find that $c_{AD}\approx 0$ (i.e.\ $c_{AB}\gg c_{AD}$), indicating that for these cases essentially all of the flux from $A$ to $E$ goes via the shorter route near saddle $D$.
Thus, in this regime, one should omit the first term in Eq.~\eqref{eq:15}.}

\begin{figure*}
    \centering
    \includegraphics[width=0.8\linewidth]{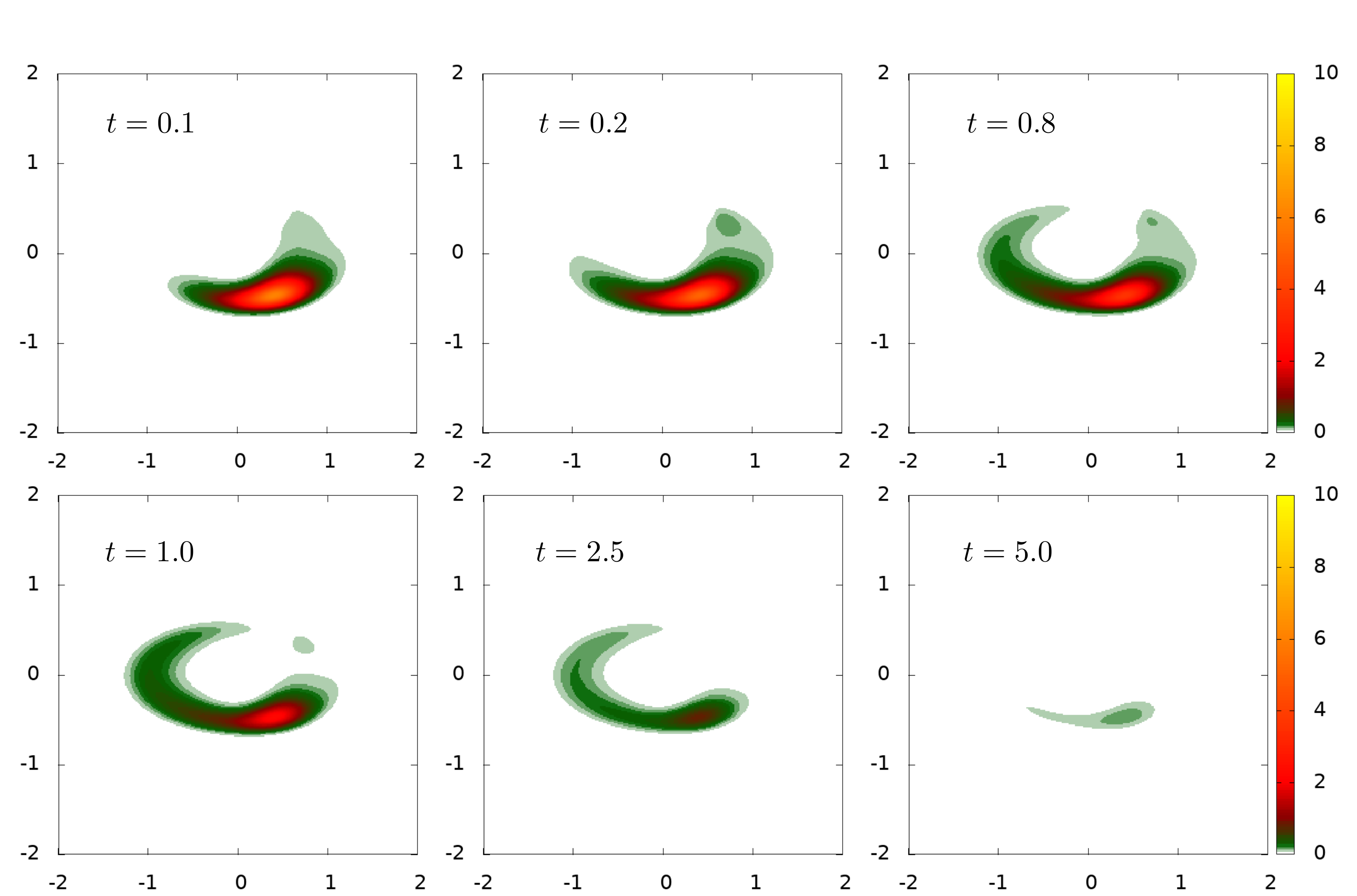}
    \caption{Time evolution of the probability density $\rho(\boldsymbol{x},t)$, for the times indicated and temperature $T^*=1$. The initial condition corresponds to the system starting a point $A$, i.e.\ with $\rho(\boldsymbol{x},t=0)=\delta(\boldsymbol{x}-\boldsymbol{x}_A)$.}
    \label{fig:rhos_beta1}
\end{figure*}

\begin{figure}[t!]
    \centering
    \includegraphics[width=0.49\linewidth]{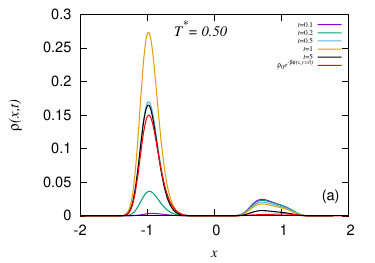}    
    \includegraphics[width=0.49\linewidth]{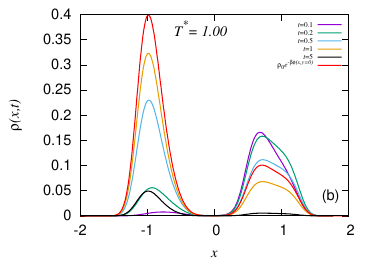}
    \includegraphics[width=0.49\linewidth]{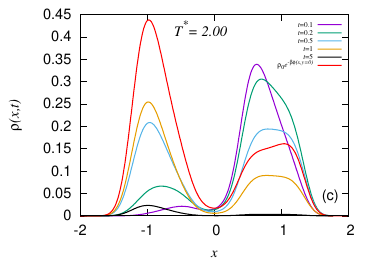}
    \caption{Time evolution of the probability density along the $x$-axis, $\rho(x,y=0,t)$, for the times indicated. The initial condition corresponds to the system starting at point $A$, i.e.\ $\rho(\boldsymbol{x},t=0)=\delta(\boldsymbol{x}-\boldsymbol{x}_A)$.}
    \label{fig:rhos_slice}
\end{figure}

Returning to Fig.~\ref{fig:seg}, there are several other features worth commenting on. The first is to note the deviations from the exponential form \eqref{eq:exp_distrib} at short times. These deviations are barely visible on the time scales of the low temperature plot in Fig.~\ref{fig:seg}(a), but for the other temperatures in Fig.~\ref{fig:seg}(b)--(d), we clearly see that for very short times, the arrival time probability is $\approx0$. This is because very fast travel directly to the destination, albeit possible, is highly unlikely.

Another feature of the plots in Fig.~\ref{fig:seg}, particularly notable in panels (b) and (c), is that the arrival time probability $p(t)$ for $x_c>0$ is much higher at short times than the corresponding $p(t)$ for $x_c<0$. This is a feature of the data for all temperatures, but for the scale of the plot displayed in Fig.~\ref{fig:seg}(a) it is not really visible. This shows that at short times the probability of traveling to the destination $E$ over the shorter distance path with the higher barrier at $D$ is much higher than via the longer path via $B$.

In Fig.~\ref{fig:log_t_x_cross} we plot the histogram of the $x$-axis crossing point $x_c$ versus the logarithm of the arrival time, $\ln(t)$, for a total of $2\times10^6$ different independent runs. This figure also shows clearly that at early times the most likely route from point $A$ to point $E$ is via the shorter route with $x_c>0$. It also shows that at lower temperatures, going near $x_c=0$ is highly unlikely, whilst at higher temperatures like $T^*=2$, this occurs fairly regularly.

The histograms in Fig.~\ref{fig:log_t_x_cross} also show that as time increases, the probability of going via either of the two routes drops off at the same (log) time.
This can be seen from the fact that the two high-density `cloud' regions in each plot end at the same time. As mentioned earlier in our discussion of Eq.~\eqref{eq:15}, this shows that at long times there is a correlation between the long-time probabilities for going to the destination via the two different paths, albeit as the good fit to Eq.~\eqref{eq:15} in Fig.~\ref{fig:tau_tau-1} shows, this can only be a weak correlation.
To be specific, we mean that these two processes are interrelated, because they depend on the common dynamical processes within the potential well at $A$.
This correlation stems from the particles that survive for a long time in the potential well at $A$ being able to repeatedly attempt to traverse each of the two barriers, before finally succeeding over one of them.
In other words, in the long time limit, the time taken to get started on the journey to cross either of the barriers, is much longer than the actual time it takes to cross over either of the barriers.

\section{Time evolution of the probability density}
\label{sec:6}

\begin{figure}[t!]
    \centering
    \includegraphics[width=0.49\linewidth]{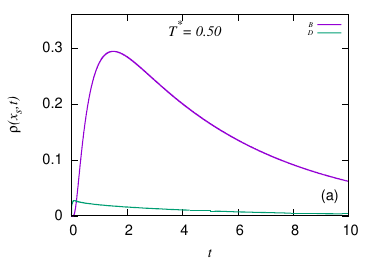}
    \includegraphics[width=0.49\linewidth]{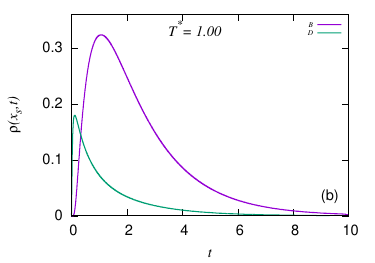}
    \includegraphics[width=0.49\linewidth]{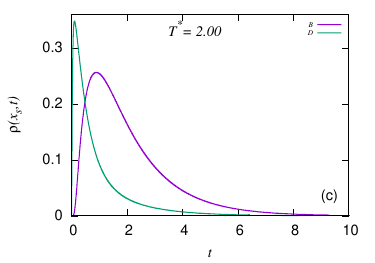}
    \caption{Density at the two saddle points $B$ and $D$ over time. The initial condition corresponds to the system being initiated at point $A$, i.e.\ $\rho(\boldsymbol{x},t=0)=\delta(\boldsymbol{x}-\boldsymbol{x}_A)$.}
    \label{fig:rhos_saddles}
\end{figure}

\begin{figure}
    \centering
    \includegraphics[width=0.49\linewidth]{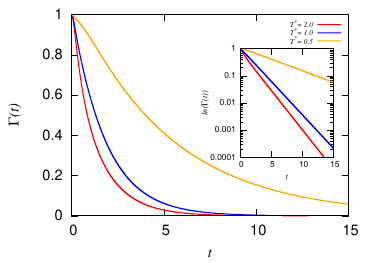}
    \caption{Total probability as a function of time, for the system to still be seeking the destination, given by Eq.~\eqref{eq:int_rho}, for three different temperatures, as indicated. In the inset we plot the same data using semi-logarithmic axes, to highlight the exponential decay at intermediate and large times.}
    \label{fig:total_density}
\end{figure}

We now move on to present results from solving numerically the Fokker--Planck equation \eqref{eq:Smolu}.
We use the finite difference algorithm developed in Ref.~\cite{chalmers2017dynamical}, on a square grid, with spacing $\Delta x=\Delta y = 0.015$, and with time-step $\Delta t=10^{-6}$.
Going beyond the approach used in Ref.~\cite{fitzgerald2023stochastic}, here we have absorbing boundary conditions surrounding the destination point $E$.

Recall that in the BD simulations presented above in Secs.~\ref{sec:4} and \ref{sec:5}, we initiate the system at point $A$. This corresponds to the initial density being $\rho(\boldsymbol{x},t=0)=\delta(\boldsymbol{x}-\boldsymbol{x}_A)$.
However, for use with any finite difference algorithm, the initial profile must be continuous and sufficiently slowly varying in space.
Therefore, we assume that at very early times $t<0.008$, the density evolves as if $\phi=0$ everywhere, which gives $\rho(\boldsymbol{x},t)$ to be the usual spreading Gaussian shape \cite{gardiner1985handbook}.
We then take the corresponding free diffusion density at the small time $t=0.008$ as our initial condition and subsequently evolve numerically under the influence of the potential $\phi$.
Note that our results for subsequent times are not sensitive to the precise value of this starting time.
One might also expect this from estimating the time for local equilibration in the minimum at point $A$ to occur.
This can be done using the Ornstein--Uhlenbeck result for the time evolution of the density in a parabolic potential \cite{archer2025timescales}.
From this result one sees that local equilibration occurs for times sufficiently large that the quantity $\exp(-2\lambda t/\gamma)\ll 1$, where $\lambda$ is the smallest eigenvalue of the Hessian of the potential at point $A$.
For $T^*=1$ we have $\lambda\approx16$, so in this case we estimate local equilibration occurs at times $t\sim0.1$.
This can also be inferred from Fig.~\ref{fig:rhos_beta1}, which shows the time evolution of the density $\rho(\boldsymbol{x},t)$.
This local equilibration time estimate also demonstrates why our results are insensitive to starting the particle exactly at point $A$, as opposed to somewhere in the vicinity of $A$, as long as the alternative start point is well within the potential well at $A$.

Recall also that in our BD simulations we define the system as having reached its destination when the distance from point $E$, $|\boldsymbol{x}-\boldsymbol{x}_E|<0.1$, and at that point it is then stopped.
Or, equivalently, it is removed from the system.
Thus, the corresponding boundary condition on Eq.~\eqref{eq:Smolu} is to set $\rho(\boldsymbol{x},t)=0$ for all $|\boldsymbol{x}-\boldsymbol{x}_E|<0.1$, $t>0$. 
It is important to mention that because of this absorbing boundary condition, the density $\rho(\boldsymbol{x},t)$ is the probability density over points in state space $|\boldsymbol{x}-\boldsymbol{x}_E|>0.1$.
Owing to the fact that we stop the system when it reaches the destination, the density $\rho(\boldsymbol{x},t)$ is not normalised over time, i.e.\ the quantity
\begin{equation}
    \Gamma(t)=\int_{-\infty}^\infty\int_{-\infty}^\infty\rho(\boldsymbol{x},t) dy \, dx,
    \label{eq:int_rho}
\end{equation}
decreases over time, with $\Gamma(t=0)=1$ and $\Gamma(t>0)<1$.
Correspondingly, the probability at time $t$ for the system to have reached the destination and to have been removed at some time previously, is $[1-\Gamma(t)]$. This is the cumulative density for the first passage process, which has probability density function
\begin{equation}
p(t) = -\frac{{\rm d}\Gamma}{{\rm d}t} = -{{\cal D}}\oint_{\partial V}\nabla\rho\cdot d\bm{S},
\label{eq:18}
\end{equation}
i.e.\ the flux over the absorbing boundary $\partial V$ surrounding point $E$. This follows from Eq.~\eqref{eq:Smolu} and the condition $\rho = 0$ on the absorbing boundary.
We do not use Eq.~\eqref{eq:18} to calculate to probability density $p(t)$, using instead BD simulations to obtain the results displayed in Fig.~\ref{fig:seg} [see also the approximate Eq.~\eqref{eq:exp_distrib}], since the BD simulation method used in Sec.~\ref{sec:5} also allows us to split the contributions to $p(t)$ according to the route taken to the destination.

In Fig.~\ref{fig:rhos_beta1} we present typical results for the time evolution of the density $\rho(\boldsymbol{x},t)$, for the temperature $T^*=1$.
It is instructive to compare these results with those in Ref.~\cite{fitzgerald2023stochastic}, which were derived with no absorbing boundary conditions around the destination point.
We see that at the early time $t=0.1$ the density has spread to much of the region within the potential well with minimum at point $A$ and is starting to cross via the barrier around point $D$. It is not until later times $t\sim0.8$ that the probability density for being near the saddle at $B$ on the longer path becomes sizable, which can also be seen from Fig.~\ref{fig:seg}(c). As time proceeds, the probability for the system to still be evolving (not removed) $\Gamma(t)$ in Eq.~\eqref{eq:int_rho} decreases, so that by the time $t=5$, the density is fairly small everywhere.

In Fig.~\ref{fig:rhos_slice} we present cross sections along the $x$-axis through the density distribution $\rho(\boldsymbol{x},t)$, i.e., $\rho(x,y=0,t)$ for various times.
In Fig.~\ref{fig:rhos_slice}(b) we present slices corresponding to the full profiles displayed in Fig.~\ref{fig:rhos_beta1}, while in Fig.~\ref{fig:rhos_slice}(a) and (c) we present the corresponding results for two different temperatures, as indicated.
It is instructive to compare Fig.~\ref{fig:x_c_histogram} and Fig.~\ref{fig:rhos_slice}.

Figure~\ref{fig:rhos_slice} shows that at early times the density around $x\approx0.7$ is always much higher than one would expect based on the equilibrium distribution \eqref{eq:rho_eq}, which is also displayed.
This peak corresponds to the system going via the shorter route.
It is only at later times $t\sim 1$ that the other peak at $x\approx -1$ fully develops.
Subsequently, over time, both peaks shrink as the probability of the system to have not reached the destination decreases.

In Fig.~\ref{fig:rhos_saddles}, we plot the density at the two saddle points $S$ over time, where $S=\{B,D\}$.
These are not the same as the arrival time distributions in Fig.~\ref{fig:seg}, but it is instructive to compare these two sets of figures.
These exhibit a very similar exponential decay form for large $t$.
The fact that at small times the most likely path from $A$ to $E$ is via the vicinity of $D$ shows in the fact that the peak in the density at point $D$ is at a much earlier time than the maximum in the density at point $B$.

Figure~\ref{fig:total_density} shows the probability given by Eq.~\eqref{eq:int_rho}, for the system to have not yet reached the destination. We present results for three different temperatures. The inset shows the same data using semi-logarithmic axes, to highlight the exponential decay at intermediate and large times, consistent with Eq.~\eqref{eq:15}. As remarked previously, it is noteworthy that the temporal behaviour is so relatively simple, given the complex spatial barrier crossing statistics, and the competition between two separate pathways.

\section{Concluding Remarks}
\label{sec:7}

Stochastic dynamics and barrier crossing processes have significant relevance across a plethora of scientific disciplines. Here, we have presented results for a relatively simple and generic model for stochastic dynamical processes that have a choice of routes to take, starting from a local minimum (metastable state) and evolving to a final (global equilibrium) state. One route takes a relatively short distance in configuration space, but over a higher barrier, while the other longer route goes over a much smaller barrier. We have shown that in this system, the probability for going over the higher barrier is much greater than one would expect based on Eq.~\eqref{eq:VHAL} or other such estimates based solely on properties of the potential at the barriers.
We also see that at higher temperatures the path over the higher barrier is taken more often than the longer route over the lower barrier.

Our results thus show the importance of three things, which must be taken into account when determining the rates of rare processes: (\textit{i}) How the system is initiated, e.g., starting in a local minimum means that the system is intrinsically far from equilibrium, so any theory used must be base on nonequilibrium statistical mechanics and any estimates based on equilibrium assumptions should be treated cautiously.
(\textit{ii}) When there is a competition between two or more paths for the system to take, the distance to travel in state space is arguably just as important as the heights of the barriers to be surmounted.
(\textit{iii}) The time available for the barrier crossing processes is crucial.
When only a short time is available compared to the equilibration time for the whole system, then the calculation of rates needs to take this into account.
Moreover, one should certainly not use observed rates to infer barrier heights as seems to be done routinely in some areas.

{For future work, to apply the insights from the present study to more realistic systems, perhaps the first aspect to deal with is quantifying properly the distance between the various states in configuration space.
For obtaining nucleation rates in condensed matter systems, the work of Lutsko et al.\ \cite{lutsko2012dynamical, lutsko2019crystals, lutsko2024microscopic}, who discuss the relevant distance (metric), is highly relevant.
Taking Lutsko's theory, which combines classical density functional theory and fluctuating hydrodynamics and applying it e.g.\ to the system discussed in Ref.~\cite{yin2021transition}, which exhibits multiple transition pathways connecting crystalline and quasicrystalline phases, would be very interesting.
Note too that here we have focused solely on the overdamped dynamics \eqref{eq:OvD} case.
However, more generally, inertia is often important \cite{hanggi1990reaction}, and can for example lead to interesting effects when there are multiple barriers present \cite{moro1992multi}.}
{In realistic settings, it is difficult to describe the dynamics in terms of a potential, and it may prove more effective to coarse-grain the dynamical observables in discrete states connected in a graph; this is a powerful simplification that leads to the concept of  Markov networks \cite{hartich2021emergent}.}

A question worth addressing here, relating to the results in Fig.~\ref{fig:x_c_histogram}, where we compare $\rho_{Fc}(x_c)$, the BD result for the $x$-axis crossing point, with $\rho_{eq}(x)$, the simple equilibrium estimate in Eq.~\eqref{eq:epd}, is: What do we find if we instead solve the Fokker--Planck equation \eqref{eq:Smolu} as we did in Sec.~\ref{sec:6}, but with the density that is removed at the destination $E$ continuously re-inserted back at the start point $A$?
Doing this over time leads to a nonequilibrium steady state, with a constant total flux through the system.
We do not present these results because we find that along the line $y=0$ the resulting steady-state density distributions are (on the scale of Fig.~\ref{fig:x_c_histogram}) not significantly different from the much simpler estimate in Eq.~\eqref{eq:epd}. 
However, the full density distributions are overall very different, in particular at the start and end points $A$ and $E$, respectively.

The simplicity of the model potential considered here makes the analysis fairly straightforward, in particular the fact that many of the interesting points in configuration space all lie very close to the line $y=0$. However, with other more complex potentials, one could perform analysis similar to that done here, simply by a careful choice of the dividing boundary through configuration space, connecting all the saddle points and any relevant local maxima, and then determining the crossing point statistics and densities over time along these dividing boundaries. 
Relating these to the invariant manifolds discussed e.g.\ in Ref.~\cite{bartsch2012reaction} may be a fruitful future connection to pursue.

\begin{acknowledgments}
GA gratefully acknowledges Loughborough University Doctoral College UK and HEC Pakistan for funding support under ``Overseas Scholarships for PhD in Selected Fields Phase-III Batch 3''.
{SPF acknowledges support from the UK EPSRC, grant number EP/R005974/1.}
\end{acknowledgments}


%

\end{document}